\documentclass[prl,aps,reprint,amsmath,amssymb,showpacs,superscriptaddress,floatfix]{revtex4-1}

\usepackage[colorlinks,citecolor=blue,linkcolor=blue]{hyperref}
\usepackage{graphicx}
\usepackage{dcolumn}
\usepackage{bm}

\DeclareMathOperator{\sech}{sech}

\begin{document}

\title{Curved magnonic waveguides based on domain walls}

\author{Felipe Garcia-Sanchez}
\email{felipe.garcia@u-psud.fr}
\affiliation{Institut d'Electronique Fondamentale, UMR CNRS 8622, Universit{\'e} Paris-Sud, 91405 Orsay, France}
\author{Pablo Borys}
\affiliation{SUPA, School of Physics and Astronomy, University of Glasgow, Glasgow G12 8QQ, UK}
\author{R{\'e}my Soucaille}
\author{Jean-Paul Adam}
\affiliation{Institut d'Electronique Fondamentale, UMR CNRS 8622, Universit{\'e} Paris-Sud, 91405 Orsay, France}
\author{Robert L. Stamps}
\affiliation{SUPA, School of Physics and Astronomy, University of Glasgow, Glasgow G12 8QQ, UK}
\author{Joo-Von Kim}
\email{joo-von.kim@u-psud.fr}
\affiliation{Institut d'Electronique Fondamentale, UMR CNRS 8622, Universit{\'e} Paris-Sud, 91405 Orsay, France}

\date{\today}

\begin{abstract}
The channeling of spin waves with domain walls in ultrathin ferromagnetic films is demonstrated theoretically and through micromagnetics simulations. It is shown that propagating excitations localized to the wall, which appear in the frequency gap of bulk spin wave modes, can be guided effectively in curved geometries and can propagate in close proximity to other channels with no perceptible scattering or loss in coherence. For N{\'e}el-type walls arising from an interfacial Dzyaloshinskii-Moriya interaction, the channeling is strongly nonreciprocal and group velocities can exceed 1 km/s in the long wavelength limit for certain propagation directions.
\end{abstract}


\maketitle


Spin waves are elementary excitations of magnetically ordered systems and their currents carry spin angular momentum like electron spin currents. Because of this, spin wave dynamics in thin ferromagnetic films have gained renewed interest in the context of low-power nanoelectronics, where spin waves have been proposed as useful vectors for information transfer and processing that minimize dissipation related to charge currents~\cite{Khitun:2010dx}. Such efforts involving the generation and control of spin wave propagation, band structure design through the use of metamaterials, and interactions with spin-polarized currents form the core of magnonics~\cite{Kruglyak:2010cy, Serga:2010cw, Lenk:2011el, Demokritov:2013we}. Because magnetization dynamics is inherently nonlinear, a quantitative description of spin wave dynamics in nanostructured materials remains a challenging problem for both fundamental and applied aspects.

Before magnonic devices can become a feasible alternative for information technologies, a number of physical issues require further consideration. First, the capacity to propagate spin waves efficiently along curved paths is essential for any form of circuit design and is a crucial element of wave processing schemes that rely on spin wave interference~\cite{Hertel:2004df}. While propagation along curved wires has been demonstrated experimentally with the assistance of current-induced Oersted fields~\cite{Vogt:2012ct} that minimizes scattering~\cite{Tkachenko:2012kb},  it remains unclear whether such schemes are feasible in complex magnonic circuits. Second, wave packet dispersion can be problematic for maintaining coherence over distances of several microns, since ferromagnetic spin waves are largely dispersive, particularly at shorter wavelengths at which the exchange interaction dominates. Finally, issues related to lithography and nanofabrication, such as edge roughness or variability in device dimensions, may become prohibitive for reproducible spin wave properties at sizes below 100 nm.

Here, we present a paradigm for spin wave propagation that relies on magnetic domain walls as natural waveguides. It is well established that spin waves propagating across a domain wall experience a scattering potential, which for static Bloch-type walls is reflectionless~\cite{Winter:1961hw, Braun:1994ff} and only leads to phase shifts~\cite{Hertel:2004df, Bayer:2005ev} but can result in momentum transfer for dynamic walls~\cite{Zvezdin:1984tb, Bouzidi:1990it, LeMaho:2009ee}. Phenomena related to the latter have motivated studies examining how domain wall motion can be effected by spin waves alone~\cite{Mikhailov:1984tt, Hinzke:2011gp, Yan:2011he, Wang:2012de, Tveten:2014hb}. Instead, we focus on a class of eigenmodes localized to the domain wall center but which propagate freely in the direction parallel to the wall as a result of a confining potential. We show theoretically and through micromagnetics simulations that such modes can be channeled effectively even in curved geometries with no measurable scattering or loss of coherence, particularly for excitation frequencies in the gap of bulk spin wave modes. Furthermore, we illustrate the effects of nonreciprocity for N{\'e}el-type domain walls resulting from a finite Dzyaloshinskii-Moriya interaction, which leads to weaker wave packet dispersion for certain propagation directions.

\begin{figure}
\includegraphics[width=8.5cm]{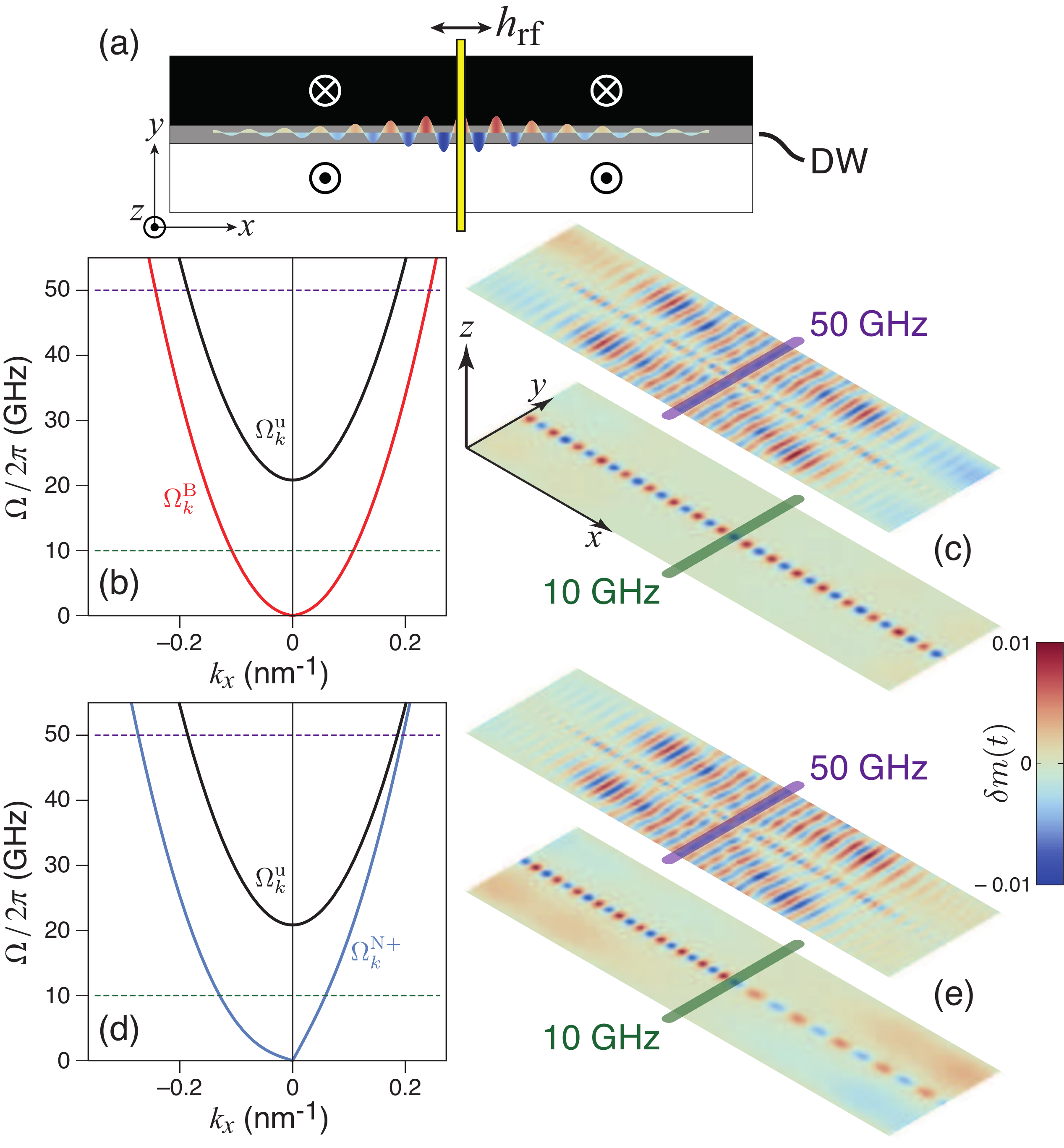}
\caption{(Color online) A magnonic waveguide based on a domain wall (DW). (a) Geometry for spin wave propagation along the center of the wall, where an radiofrequency antenna generating an alternating field $h_{\rm rf}$ excites spin waves that propagate along the ($x$). (b, d) Dispersion relation for channeled Bloch (b, red curve), $\Omega_k^{\rm B}$, and N{\'e}el (d, blue curve), $\Omega_k^{\rm N}$, domain wall spin wave modes in comparison with bulk spin waves (black curve), $\Omega_k^{\rm u}$. For the N{\'e}el wall case, $D = 1.5$ mJ/m$^2$. (c, e) Simulation results of propagating modes for excitation field frequencies in the bulk (50 GHz) and in the gap (10 GHz) for Bloch (c) and N{\'e}el (e) walls. These driving frequencies are shown as dashed lines in (b, d).}
\label{fig:geometry}
\end{figure}

The basic principle underpinning the domain wall magnonic waveguide is illustrated in Fig.~\ref{fig:geometry}.
Consider a thin rectangular ferromagnetic wire with perpendicular magnetic anisotropy along the $z$-axis. A domain wall separates two uniformly magnetized ``up'' and ``down'' states with the wall axis along $y$, which is perpendicular to the wire axis $x$. The spin waves considered are associated with localized domain wall eigenmodes that propagate along the $x$ direction, parallel to the domain wall. In the micromagnetics simulations used~\cite{Vansteenkiste:2014et}, these modes are driven by a microwave antenna that is modeled as a line source of a sinusoidal excitation field $h_{\rm rf}$.

In the presence of an isotropic exchange and dipole-dipole interactions, the Bloch-type domain wall minimizes the volume dipolar interaction and it is characterized by moments that rotate in a plane ($xz$) perpendicular to the wall direction ($y$). For this wall type, it has been shown by Winter that there exists a family of spin wave eigenmodes,
\begin{equation}
\psi_k(x,y,t) = \exp\left[ i (\Omega_k^{\rm B} t - k_x x  ) \right] \sech\left( y / \lambda \right),
\label{eq:WinterModes}
\end{equation}
which are localized in the direction perpendicular to the domain wall ($y$) on a length scale $\lambda$ but propagate as plane waves parallel to the domain wall ($x$)~\cite{Winter:1961hw}. Here, $\lambda = \sqrt{A/K_0}$ represents the characteristic wall width parameter where $A$ is the exchange, $K_0 = K_u - \mu_0 M_s^2/2$ is the effective perpendicular anisotropy constant, and $M_s$ is the saturation magnetization. These modes are exchange-dominated spin wave modes that are described by the dispersion relation
\begin{equation}
\Omega_k^{\rm B} = \sqrt{\omega_k \left( \omega_k + \omega_\perp \right)},
\label{eq:BlochDispersion}
\end{equation}
where $\omega_k = 2 \gamma A k_x^2 / M_s$ is the quadratic exchange part, $\omega_\perp = 2 \gamma K_\perp / M_s$, and $K_{\perp} = \mu_0 N_y M_s^2 / 2$ is a transverse anisotropy that represents the dipolar interaction due to volume charges at the domain wall center, with $N_y$ representing an effective demagnetization constant along the wall axis $y$. For ultrathin films $N_y \approx d / (d + \lambda)$, where $d$ is the film thickness~\cite{Mougin:2007hz}. Note that these modes are gapless because the effective field associated with the perpendicular anisotropy cancels out at the wall center. In Fig.~\ref{fig:geometry}(b), the dispersion relation (\ref{eq:BlochDispersion}) is shown in comparison with the usual bulk spin wave modes for the uniformly magnetized state,
\begin{equation}
\Omega_k^{\rm u} = \omega_k + \frac{2 \gamma K_0}{M_s},
\label{eq:BulkDispersion}
\end{equation}
where we have assumed $A = 15$ pJ/m, $K_u = 1$ MJ/m$^3$, $M_s = 1$ MA/m, and $d = 1$ nm. For a microwave field excitation in the frequency gap of the bulk modes $\Omega_k^{\rm u}$, which is determined by $2 \gamma K_0/M_s$,  one observes that only the localized Winter modes $\Omega_k^{\rm B}$ are excited and are effectively channeled along the domain wall center [Fig.~\ref{fig:geometry}(c), excitation at 10 GHz], which acts as a local potential well for these spin wave modes. When the microwave field is applied in the frequency band of the bulk modes, the channeling phenomenon is preserved whereby the localized modes can be seen to propagate with a higher wave vector than the bulk modes [Fig.~\ref{fig:geometry}(c), excitation at 50 GHz].

For ultrathin ferromagnetic films in contact with a large spin-orbit coupling material in asymmetric multilayer stacks, an interfacial Dzyaloshinskii-Moriya interaction (DMI)~\cite{Fert:1980hr, Fert:1990, Crepieux:1998ux, Bogdanov:2001hr} may appear that favors  a N{\'e}el-type domain wall profile as a ground state~\cite{Heide:2008da, Thiaville:2012ia}. The moments in this wall type rotate in a plane ($yz$) that is parallel to the wall direction ($y$), which leads to an increase in the volume dipole-dipole interaction but which is subsequently compensated by the DMI above a critical value, $D>D_c$~\cite{Thiaville:2012ia}. In this case, the inclusion of the DMI leads to a hybridization of the Winter modes~\cite{GarciaSanchez:2014dw}. Nevertheless, an expression of the channeling spin wave mode frequencies for N{\'e}el-type walls can be found using perturbation theory by using the Winter modes in (\ref{eq:WinterModes}) as a scattering basis, which involves computing frequency shifts due to terms such as $\langle \psi_k | \sech{\left(y/\lambda\right)} | \psi_k \rangle$ and $\langle \psi_k | \sech{\left(y/\lambda\right)} \partial_x | \psi_k \rangle$~\cite{GarciaSanchez:2014dw}. The resulting eigenfrequencies are found to be
\begin{equation}
\Omega_k^{\rm N\pm} = \sqrt{\omega_k \left( \omega_k - \omega_\perp + \frac{\omega_{D,k}}{k \lambda} \right)} \pm \omega_{D,k},
\label{eq:NeelDispersion}
\end{equation}
where $\omega_{D,k} = \pi \gamma D k_x / 2 M_s$. In addition to an ellipticity in the precession, the DMI results in a linear wave vector dependence for the mode frequency, which is consistent with behavior in other geometries~\cite{Udvardi:2009fm, Zakeri:2010ki, Moon:2013dm, GarciaSanchez:2014dw}. However, this linear $k_x$ dependence does not lead to a simple shift in the quadratic dispersion relation as a result of the ellipticity. Instead, the dispersion relation becomes markedly asymmetric with respect to $k_x = 0$ [Fig.~\ref{fig:geometry}(d)], where a quasi-linear variation is seen for  $k_x>0$ while a strongly quadratic variation is preserved for $k_x<0$. This asymmetry leads to pronounced differences in the left- and right-propagating wave vectors, which can be seen for microwave field excitations in the frequency gap and in the frequency band of the bulk spin wave modes [Fig.~\ref{fig:geometry}(e)]. The channeling properties of the N{\'e}el-type wall are preserved even in cases where the localized and propagating mode frequencies are closely spaced, which can be seen for the $k_x >0$ propagation at around 50 GHz in Fig.~\ref{fig:geometry}(e). Note that in the limit of $k \rightarrow 0$ and $\omega_\perp \rightarrow 0$, Eq. (\ref{eq:NeelDispersion}) predicts an instability in the domain wall ground state $\Omega_k^{\rm N-} = 0$ for a critical value of the DMI, $D_{c2} = 4\sqrt{A K_0}/\pi$, which is consistent with previous work~\cite{Heide:2008da, Thiaville:2012ia}. Our micromagnetics simulations indeed show that straight domain walls become unstable for $D > D_{c2}$.

\begin{figure}
\includegraphics[width=8.5cm]{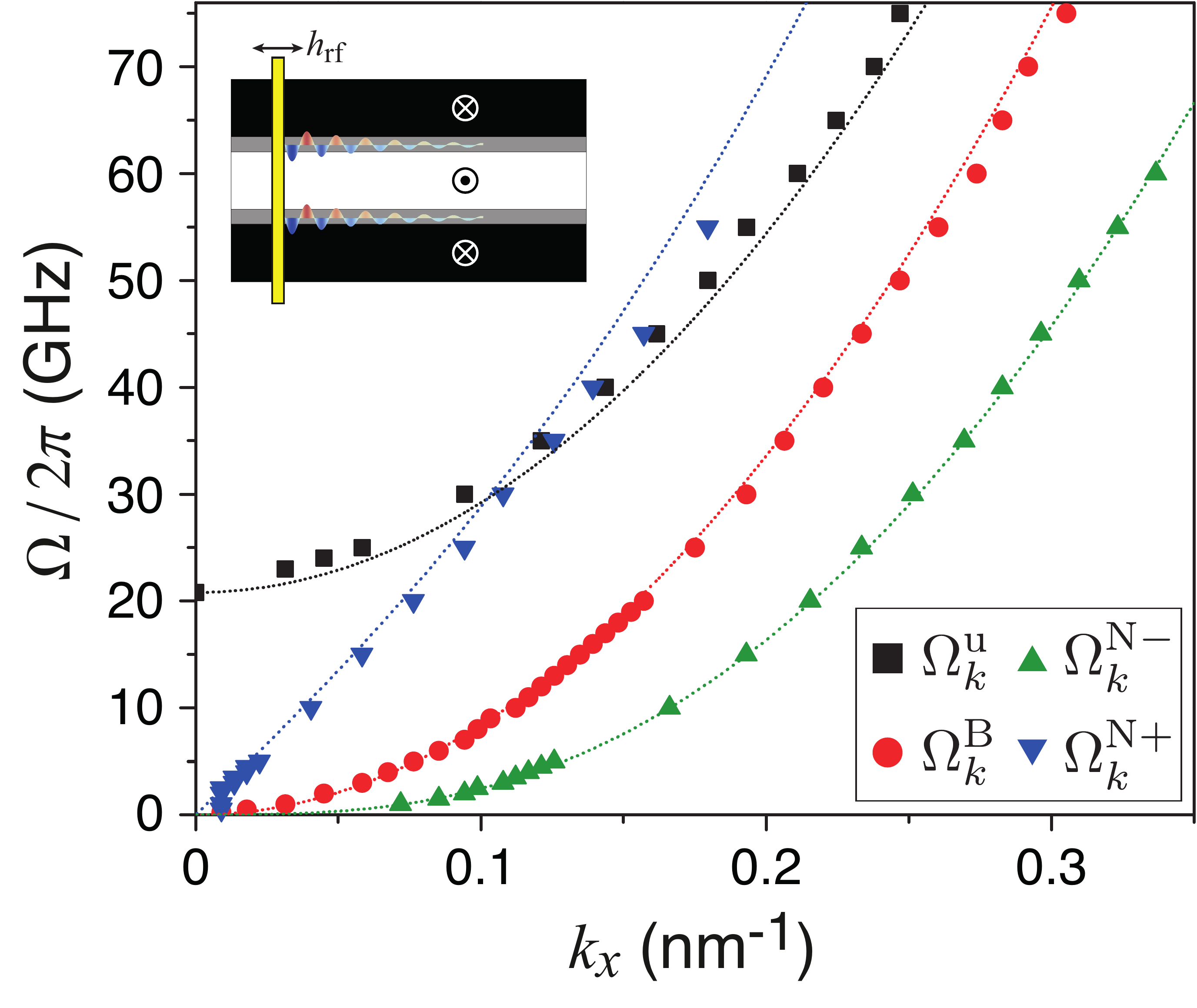}
\caption{\label{fig:dispersion-relation}(Color online) Simulated dispersion relation for channeled and bulk spin waves. $\Omega_k^{\rm u}$ indicate bulk modes. The channeled modes for Bloch-type ($\Omega_k^{\rm B}$) and N{\'e}el-type ($\Omega_k^{\rm N\pm}$) walls are computed, where $D = 3$ mJ/m$^2$ for the latter and the sign indicates propagation relative to the wall chirality. The points represent simulated values and lines are based on Eqs. (\ref{eq:BlochDispersion}), (\ref{eq:BulkDispersion}), and (\ref{eq:NeelDispersion}). The inset shows the three-domain geometry with two parallel domain wall channels.} 
\end{figure}

To determine the accuracy of the perturbation theory and to explore simultaneous propagation along multiple channels, the spin wave mode frequencies were computed using micromagnetics simulations for a three-domain structure comprising two parallel domain walls. The simulated dispersion relations for $k_x >0$ propagation in this configuration are presented in Fig.~\ref{fig:dispersion-relation}.
As before, two distinct systems are considered, Bloch-type walls ($D=0$) and N{\'e}el-type walls ($D=3$ mJ/m$^2$). For all channeled domain wall and bulk modes, the analytical theory gives a good account of the simulated dispersion relations, where discrepancies may arise due to limits in the wave vector resolution resulting from the finite size of the simulation grid. For the N{\'e}el wall case, we observe two frequency branches since the propagation in the two domain wall channels occur for different relative chiralities of the domain wall. No discernible interference is observed between the two channels, which indicates that the wall modes can propagate with distinct wave vectors along the different channels with minimal ``crosstalk''. An interesting property of these N{\'e}el wall branches is that the difference in their frequencies,  $\Delta\Omega_{k}^{\rm N}= \Omega_{k}^{\rm N+} - \Omega_{k}^{\rm N-}=2\omega_{D,k}$,  is simply proportional to the constant $D$. Therefore, a simultaneous measurement of these two branches may offer a means of probing the strength of the DMI in such systems.

\begin{figure}
\includegraphics[width=8cm]{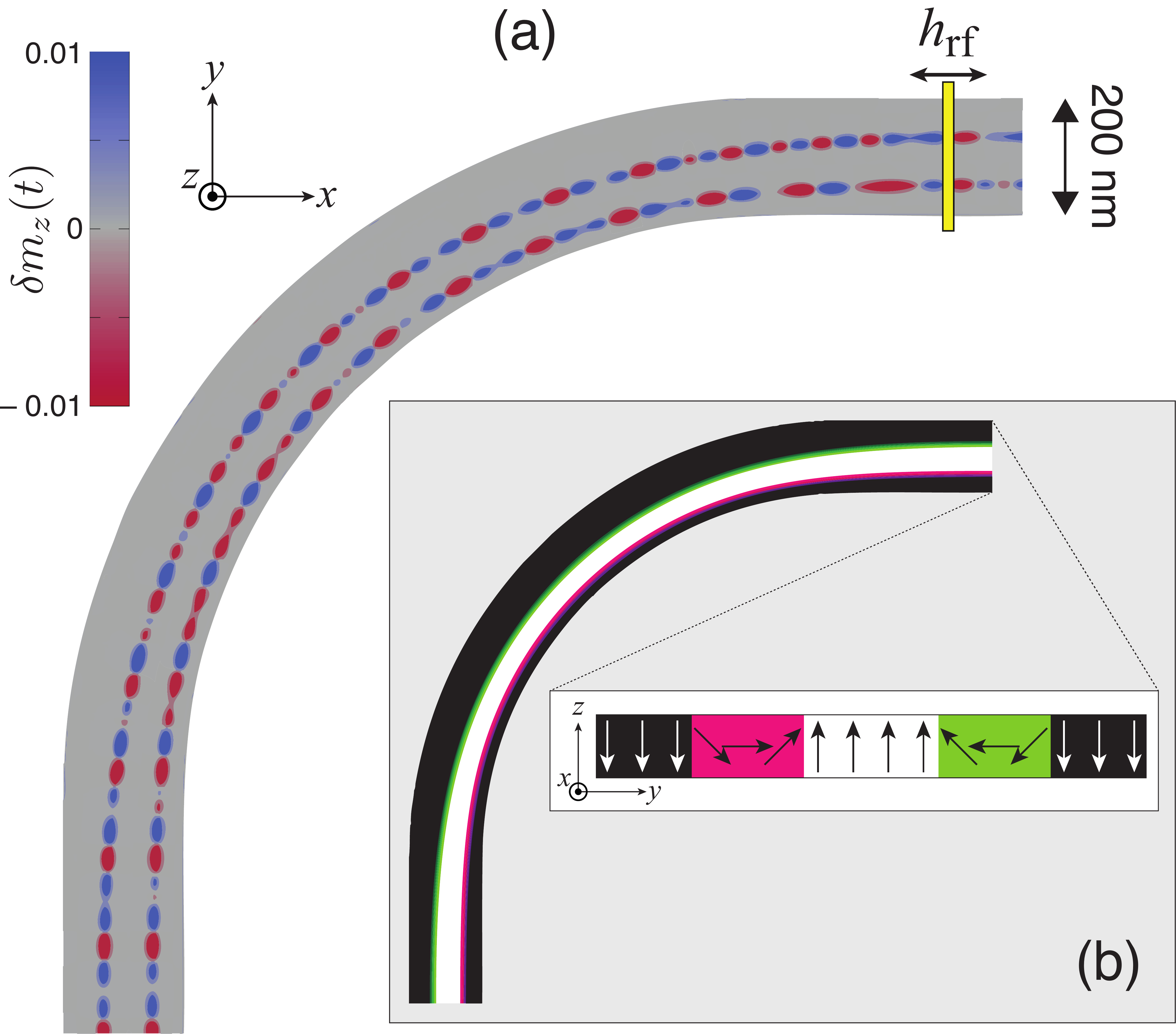}
\caption{\label{fig:curved-configuration}(Color online) (a) Spin wave channeling around a curved track through two N{\'e}el-type domain walls ($D = 1$ mJ/m$^2$). Fluctuations in the $m_z$ magnetization component are shown as a color code for a microwave excitation frequency of 5 GHz, where the simulated microwave antenna is situated at the top of the track. The area of the simulated region is 1600 nm $\times$ 1600 nm. The width of the wire is 200 nm and the thickness is 1 nm. (b) Equilibrium configuration of the curved track comprising three domains. The inset shows a schematic of the magnetization profile in a cross section at the top of the track.}
\end{figure}

The possibility of guiding spin waves using domain walls in curved geometries is shown in Fig.~\ref{fig:curved-configuration}.
A 200 nm wide curved track with a 90$^\circ$ degree bend is considered, with a radius of curvature of approximately 1600 nm for the outer edge.  The magnetic state of the track comprises a three-domain structure, where the domain walls run approximately parallel to the track edges. In order to stabilize this domain state, a DMI of $D = 1$ mJ/m$^2$ was used to ensure that the domain walls are not expelled from the track as a result of dipolar interactions. A microwave antenna is placed at one end of the track, which excites spin waves that propagate in a counterclockwise direction along the track. Fig.~\ref{fig:curved-configuration} illustrates the propagation for an excitation frequency of 5 GHz, which is in the frequency gap of the bulk modes. One observes a clear channeling effect as a result of the domain wall waveguides, where the spin waves can be seen to propagate along the curved track without any apparent scattering or loss of coherence, and again there is no perceptible interference between the two domain wall channels. The nonreciprocal effect due to the different relative chiralities seen for the propagating modes is also preserved, which suggests that closely-spaced domain walls can act as independent channels along as excitations in the frequency gap of the bulk modes are considered. While this geometry may be difficult to realize experimentally, it serves to illustrate the salient features of the domain wall magnonic waveguides in a curved geometry.

\begin{figure}
\includegraphics[width=8cm]{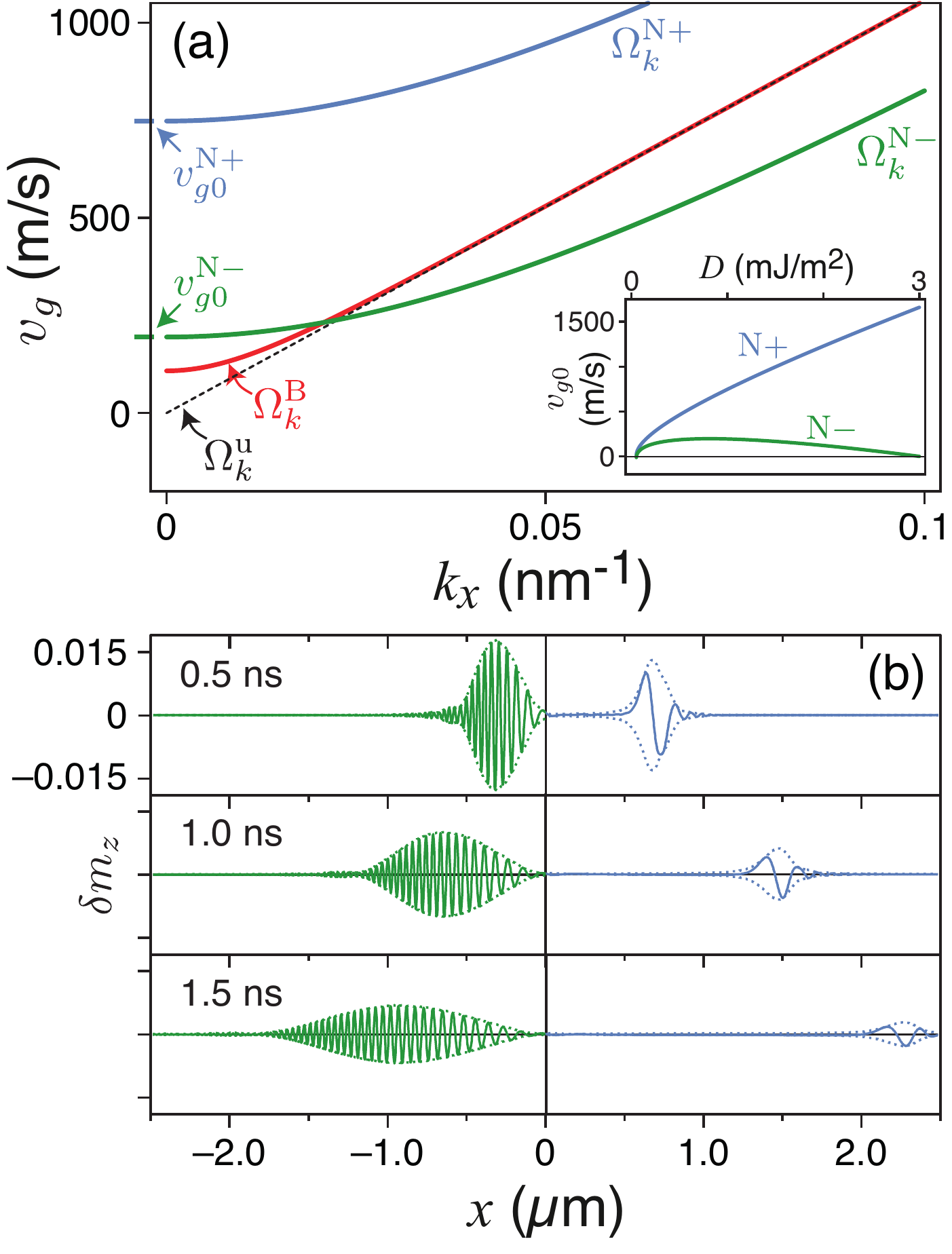}
\caption{\label{fig:wavepackets}(Color online) (a) Group velocity for the channeled and bulk spin waves.  $\Omega_k^{\rm u}$ indicate bulk modes. The channeled modes for Bloch-type ($\Omega_k^{\rm B}$) and N{\'e}el-type ($\Omega_k^{\rm N\pm}$) walls ($D = 1$ mJ/m$^2$) are computed, where the sign for the latter indicate propagation relative to the wall chirality. The inset shows the $k\rightarrow 0$ limit of the group velocity for the two N{\'e}el wall modes as a function of the Dzyaloshinskii-Moriya constant, $D$. (b) Wave packets in a N{\'e}el wall channel ($D = 3$ mJ/m$^2$) at three instants after generation by a sinusoidal pulse with $\nu_p = 7.5$ GHz.}
\end{figure}

The dispersion relation for the N{\'e}el wall modes has interesting consequences for wave packet propagation. In the ultrathin film geometry in which a strong perpendicular anisotropy is present, the bulk spin wave spectrum is mainly exchange-dominated and exhibits a quadratic dispersion relation, as illustrated in Figs.~\ref{fig:geometry}(b) and \ref{fig:dispersion-relation}. As such, wave packets comprising bulk spin waves in the long wavelength limit, $k \rightarrow 0$, exhibit strong dispersion and a vanishing group velocity, $v_g$, since $v_g^u \equiv \partial_k \Omega_k^u = 4 \gamma A k/M_s$ is linear in $k$ and vanishes as $k \rightarrow 0$. For Bloch-type walls, $v_{g}$ for the channeled mode remains finite in this limit as a result of the weak ellipticity of the precession, with a value of $v_{g0}^{\rm B} \equiv v_{g}^{\rm B}(k=0) = \gamma \sqrt{2 A K_\perp}/M_s$, which is approximately 76 m/s with the numerical parameters considered here.  For N{\'e}el-type walls, the channeled spin wave modes in the long wavelength limit possess a group velocity characterized by
\begin{equation}
v_{g0}^{\rm N\pm} = \frac{\gamma}{M_s} \left(  \sqrt{A \left( \frac{\pi D}{\lambda} - 2 K_\perp \right)}  \pm \frac{\pi D}{2}  \right),
\end{equation}
which is found from Eq. (\ref{eq:NeelDispersion}). A striking feature is that the group velocity of the two branches can have large finite values even in the long wavelength limit, but which are strongly dependent on the propagation direction as shown in Fig.~\ref{fig:wavepackets}.
The difference in $v_{g0}^{\rm N\pm}$ between the two branches is simply proportional to $D$, but the functional form of $v_{g0}^{\rm N\pm}(D)$ itself is nontrivial and is shown in the inset of Fig.~\ref{fig:wavepackets}(a). It is interesting to note that the group velocity for the $\Omega_k^{\rm N\-}$ branch tends towards zero as $D$ increases, which is mirrored by an increasing group velocity for the $\Omega_k^{\rm N+}$ mode. For a moderate value $D= 1.5$ mJ/m$^2$, $v_{g0}^{\rm N+} \approx $ 1000 m/s, which might be a useful characteristic for information technologies.

An illustration of nonreciprocal wave packet propagation computed using micromagnetics simulations is shown in Fig.~\ref{fig:wavepackets}(b). The wave packets are generated with a field pulse at $x=0$ that comprises a sine wave oscillation, $h_p = h_{p0} \sin(2 \pi \nu_p t)$ with $h_{p0} = $ 100 mT and $\nu_p = 7.5$ GHz, over one period. This form allows better wave vector selection in reciprocal space, since the Fourier transform of this function is peaked at a finite value of $k_x$. The wave packet is generated in a straight N{\'e}el-type wall [\emph{cf}. Fig.~\ref{fig:geometry}(e)] with a large value of the DMI, $D = 3$ mJ/m$^2$, in order to better visualize the nonreciprocity. The temporal evolution of the $m_z$ component of the wave packet is shown for three instants after the initiation of the pulsed field. Propagation towards the $-x$ direction involves the $\Omega_k^{\rm N-}$ branch and exhibits strong dispersion, where the wave packet spreads out over a micron after 1.5 ns. On the other hand, the propagation along the $+x$ direction ($\Omega_k^{\rm N+}$) exhibits a much weaker dispersion where the wave packet can be observed to retain its shape after propagating over 2 $\mu$m. The decrease in amplitude for the $\Omega_k^{\rm N+}$ packet is related to Gilbert damping ($\alpha = 0.01$), rather than dispersion. The wave packet velocity for $\Omega_k^{\rm N+}$ computed from simulation is approximately 1550 m/s, which is close to the value of $\approx$ 1700 m/s expected from the analytical theory. Good agreement for the velocity is also found for the $\Omega_k^{\rm N-}$ branch, where simulations give a value of $\approx$ 750 m/s while the analytical theory predicts a group velocity of $\approx$ 775 m/s.

The channeling properties of domain walls overcome some of the issues highlighted in the introduction. Because domain walls are topological solitons their spin structure is primarily governed by intrinsic properties of the magnetic material, which makes them less sensitive to issues related to lithography or nanofabrication such as edge roughness or sample-to-sample reproducibility. Channeling along curved walls might prove to be useful for modulating propagation lengths for applications involving interference, but it would at the very least relax constraints on how straight spin wave conduits need to be for magnonic circuits  based on such waveguides to function. Finally, reconfigurable waveguide schemes based on domain structures could be envisaged, where the number, spacing, and shape of domain wall arrays could be modified with applied fields or spin-polarized currents.


\begin{acknowledgments}
The authors acknowledge fruitful discussions with M. Bailleul, R. Hertel, T. Devolder, and S. Petit-Watelot. This work was partially supported by the French National Research Agency (ANR) under contract no. ANR-11-BS10-0003 (NanoSWITI), the University of Glasgow, EPSRC, and the National Council of Science and Technology of Mexico (CONACyT).
\end{acknowledgments}

\bibliography{articles}

\end{document}